# Tensor Mixed Effects Model with Application to Nanomanufacturing Inspection

Xiaowei Yue[1], Jin Gyu Park[2], Zhiyong Liang[2], Jianjun Shi[3*]

[1] Grado Department of Industrial and Systems Engineering, Virginia Polytechnic Institute and State University, Blacksburg, VA 24061

[2] High-Performance Materials Institute, Florida State University, Tallahassee, FL 32310

[3] H. Milton Stewart School of Industrial and Systems Engineering, Georgia Institute of Technology, Atlanta, GA 30332

**ABSTRACT**

Raman mapping technique has been used to perform in-line quality inspections of nanomanufacturing processes. In such an application, massive high-dimensional Raman mapping data with mixed effects is generated. In general, fixed effects and random effects in the multi-array Raman data are associated with different quality characteristics such as fabrication consistency, uniformity, defects, et al. The existing tensor decomposition methods cannot separate mixed effects, and existing mixed effects model can only handle matrix data but not high-dimensional multi-array data. In this paper, we propose a tensor mixed effects (TME) model to analyze massive high-dimensional Raman mapping data with complex structure. The proposed TME model can (i) separate fixed effects and random effects in a tensor domain; (ii) explore the correlations along different dimensions; and (iii) realize efficient parameter estimation by a proposed iterative double Flip-Flop algorithm. We also investigate the properties of the TME model, existence and identifiability of parameter estimation. The numerical analysis demonstrates the efficiency and accuracy of the parameter estimation in the TME model. Convergence and asymptotic properties are discussed in the simulation and surrogate data analysis. The case study shows an application of the TME model in quantifying the influence of alignment on carbon nanotubes buckypaper. Moreover, the TME model can be applied to provide potential solutions for a family of tensor data analytics problems with mixed effects.

[*] Xiaowei Yue is an assistant professor at the Virginia Polytechnic Institute and State University, Blacksburg, VA 24060, USA (Email: xwy@vt.edu). Jin Gyu Park is the research faculty at the High-Performance Materials Institute, Florida State University, Tallahassee, FL 32310, USA (Email: jgpark@fsu.edu). Zhiyong Liang is the Professor and Director of the High-Performance Materials Institute, at the Florida State University, Tallahassee, FL 32310, USA (Email: liang@eng.fsu.edu). Jianjun Shi is the Carolyn J. Stewart Chair and Professor at the Georgia Institute of Technology, Atlanta, GA 30332, USA (Email: jianjun.shi@isye.gatech.edu). Dr. Shi is the corresponding author. This research is partially funded by the National Science Foundation (NSF) Scalable Nanomanufacturing Program (SNM 1344672).

**1. Introduction**

Carbon nanotubes (CNTs) buckypaper is an important multifunctional platform material with great potential for creating lightweight and high-performance materials for various applications due to buckypaper's superior mechanical and electrical characteristics. One of the critical bottlenecks in the massive production and applications of high-quality buckypaper is quality inspection and monitoring of nanomanufacturing processes. The challenges include: (i) applying quick and accurate quality metrology to obtain information associated with microstructure, (ii) characterizing and analyzing in-line data to extract useful quality information for inspection and monitoring.

As an effective characterization method for nanostructure information, Raman spectroscopy is very suitable for in-line quality inspection of nanomanufacturing processes. As an example, one Raman spectrum of single-walled CNTs buckypaper is shown in Fig. 1. In the figure, the Raman peak intensity corresponds to material concentration and distribution; peak frequency is associated with molecular structure and phase; bandwidth is associated with crystallinity and phase (Salzer and Siesler 2009); intensity ratio of D-band and G-band can be affected by degree of functionalization (Cheng et al. 2009). Therefore, numerous vital information about buckypaper quality is hidden in the Raman spectra data, which provides unprecedented opportunities for quality inspection, system informatics, and monitoring.

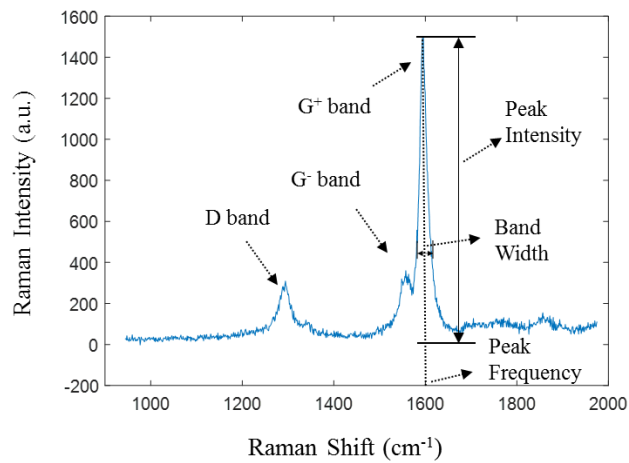

Fig. 1. One Raman Spectrum for Single-Walled CNTs Buckypaper

Due to the recent development of metrology technologies, Raman mapping (also called Raman spectral imaging) can be used to perform in-line quality inspection in continuous CNTs buckypaper nanomanufacturing processes. Raman mapping is a technique for generating detailed multi-array Raman spectra including numerous information about nanomaterials. Meanwhile, it is a challenging task to conduct data analytics, feature extraction, pattern recognition and in-line decision making, due to the high-dimensionality, large data size, as well as complex spatial and temporal correlation structures of Raman mapping. Specifically, in a Raman mapping measurement for single-walled CNTs buckypaper, about 600 Raman spectra can be collected per minute from a rectangular zone with a dimension of 10 micrometers by 60 micrometers. As shown in Fig. 2, multiple measurement points are chosen from a rectangular zone, and each measurement point generates one Raman spectrum. Every Raman spectrum includes 1024 Raman shifts and intensities. The correlations along x/y directions of the rectangular zone are different due to the alignment of carbon nanotubes in the CNTs buckypaper. Meanwhile, the correlation along the Raman shift is different from the aforementioned spatial correlation.

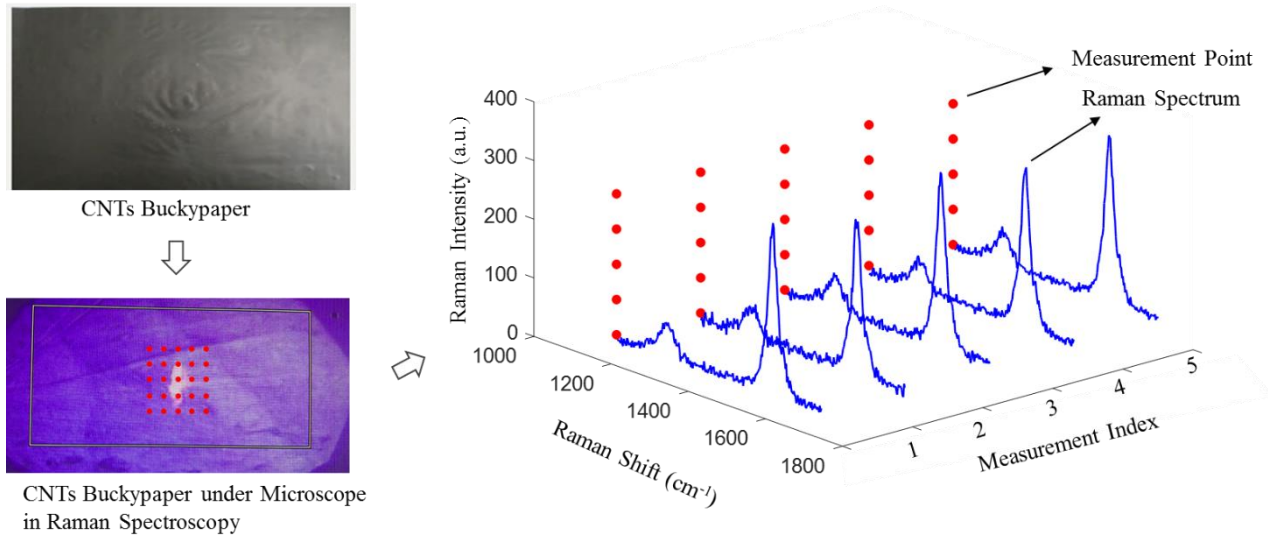

Fig. 2. Raman mapping from the rectangular zone in CNTs buckypaper

According to the data structure of Raman mapping, tensor is an efficient mathematical tool for formulating the Raman mapping for nanomanufacturing inspection. Tensors (also called multidimensional arrays) have become increasingly important because they provide a concise mathematical framework for formulating the high-dimensional data. Similar to the linear regression model in classical statistics, people use high-order tensor decompositions in high-dimensional statistics, such as CANDECOM/PARAFAC (CP) decomposition and Tucker decomposition, to separate different components inherent to the data. Kolda and Bader (2009) provided an overview of higher-order tensor decompositions, their applications, and available software. Corresponding to the generalized linear model (GLM) in statistics, Zhou et al. (2013) proposed a GLM model in the tensor domain, which extends the classical vector-valued covariate regression to an array-valued covariate regression. However, these tensor-based methods do not consider the multilevel variabilities (mixed effects) in the datasets.

A mixed effects model is a statistical model containing both fixed effects and random effects. It has nice properties, including (i) the capability to handle multilevel hierarchical data, such as longitudinal data with multiple measurements collected over time for an individual sensor; (ii) its ability to take complex association structures, including the correlation between different groups and correlation within an individual group, into consideration. Thus, the mixed effects model is widely used in a variety of disciplines such as physics, biology, engineering and social sciences (Demidenko 2013; Galecki and Burzykowski 2013). However, the classical mixed effects model treats multivariate data as a vector or a matrix, which is insufficient for analysis of high-dimensional data, such as tensor-type Raman mapping with its high dimensionality and complex correlations. Thus we need to develop a novel tensor mixed effects (TME) model that can explore mixed effects in the tensor domain.

We emphasize the motivation of developing a TME model with an example in nanomanufacturing. Raman mapping data are collected to inspect the quality of continuously fabricated CNTs buckypaper. There are multiple components in the data that are associated with different critical quality characteristics. Specifically, fixed effects measure the fabrication consistency of quality features (such as degree of alignment, degree of functionalization, nanotube distribution, and dispersion). This indicates whether there is a gradual mean shift in the roll-to-roll fabrication process of CNTs buckypaper. In addition, random effects are relevant to the uniformity and defect information. The uniformity pertains to the status of the quality indices, while the defect information consists of the number and the pattern of defects in the CNTs buckypaper (Yue et al. 2018). Therefore, it is necessary to use a mixed effects model to decompose different effects in the Raman mapping data. From another point of view, Raman mapping data have tensor structures. One Raman mapping dataset usually contains multiple dimensions: two measurement coordinates, Raman shift (frequency) and Raman intensity. If matricization or vectorization is conducted to process the Raman data, a classical mixed effects model can be developed. However, this vectorized linear mixed effects (vLME) model has three limitations: (i) the dimension after vectorization becomes very high and a large sample size is required for accurate parameter estimation; Also, vectorization destroys the tensor structure and results that the corresponding basis matrices are not full rank; (ii) the computation cost will be large, and it cannot meet the needs of in-line inspection and monitoring; (iii) the transformation alters the inherent multi-way correlation structures, which makes the correlation along different dimensions unobtainable. To overcome these three limitations, this paper proposes the tensor mixed effects (TME) model.

The TME model can effectively and efficiently explore multilevel variabilities (including fixed effects and random effects) inherent to tensor-structured high-dimensional data. It can be regarded as a logical extension from a vector-valued or matrix-valued mixed effects model to an array-valued mixed effects model. It is a challenging task to develop the TME model because (i) it deals with high-dimensional datasets with tensor structure; (ii) an efficient algorithm is required to do parameter estimation; (iii) it is necessary to ensure the identifiability of multi-dimensional correlations. In this paper, we propose the TME model and explore its properties. An iterative Flip-Flop algorithm is proposed for parameter estimation.

The remainder of this paper is organized as follows. Section 2 introduces basic tensor notation and preliminaries, and further proposes the TME model. Section 3 describes the proposed maximum likelihood estimation (MLE) algorithm for the TME model, in addition to investigating the existence of the MLE and the identifiability of the TME model. In Section 4, a double Flip-Flop algorithm is proposed to conduct parameter estimation of the TME model. In addition, initialization and convergence criteria of the algorithm are provided. Sections 5 presents a numerical simulation, a surrogated data analysis and a real case study of Raman mapping to test the performance of the TME model. Finally, a brief summary is provided in Section 6.

## 2. Tensor Mixed Effects Model

In this section, we first introduce the tensor notation and preliminaries. Then, we propose the TME model and explore the random distribution of tensor responses. Next, we discuss the maximum likelihood estimation (MLE) for the TME model, the conditions for the existence of the MLE, and the constraints to ensure the identifiability.

## 2.1 Tensor Notation and Preliminaries

In this section, basic notations, definitions, and matrix/array operators in multilinear (tensor) algebra are introduced and summarized. The terminology used here remains as consistent as possible with the terminology of previous publications (Kolda and Bader 2009; Zhou et al. 2013) in the area of tensor algebra. Scalars are denoted by lowercase italic letters, e.g., $a$; vectors by lowercase italic boldface letters, e.g., $\boldsymbol{a}$; matrices by uppercase italic boldface letters, e.g., $\boldsymbol{A}$; and tensors by calligraphic letters, e.g., $\mathcal{X}$. The order of a tensor is the number of dimensions (modes). For example, an $K$-order tensor is denoted by $\mathcal{X} \in \mathbb{R}^{I_1 \times \cdots \times I_K}$, where $I_k$ denotes the $k$-mode dimension of $\mathcal{X}$. The $i^{\text{th}}$ entry of a vector $\boldsymbol{a}$ is denoted by $a_i$, the element $(i,j)$ of a matrix $\boldsymbol{A}$ is denoted by $a_{ij}$, and the element $(i,j,k)$ of a third-order tensor $\mathcal{X}$ is denoted by $x_{ijk}$. Indices range from 1 to their capital versions, e.g., $i = 1,\cdots,I$.

Matricization, also known as unfolding or flattening, is the process of reordering the elements of a tensor into a matrix (Kolda and Bader 2009). The $k$-mode matricization of a tensor $\mathcal{X} \in \mathbb{R}^{I_1 \times \cdots \times I_K}$ is denoted by $\boldsymbol{X}_{(k)}$. $\text{vec}(\mathcal{X})$ is the vectorization of a tensor $\mathcal{X}$. The $k$-mode product of a tensor $\mathcal{X} \in \mathbb{R}^{I_1 \times \cdots \times I_K}$ with a matrix $\boldsymbol{U} \in \mathbb{R}^{J \times I_k}$ is denoted by $\mathcal{X} \times_k \boldsymbol{U}$ and elementwise, we have $(\mathcal{X} \times_k \boldsymbol{U})_{i_1 \cdots i_{k-1} j i_{k+1} \cdots i_K} = \sum_{i_k=1}^{I_k} x_{i_1 \cdots i_k \cdots i_K} u_{j i_k}$, where all the indices range from 1 to their capital versions, e.g., the index $j$ goes from $1,2,\ldots,J$, and the index $i_k$ goes from $1,2,\ldots,I_k$. The kronecker product of matrices $\boldsymbol{A}$ and $\boldsymbol{B}$ are denoted by $\boldsymbol{A} \otimes \boldsymbol{B}$. The kronecker product is an operation on two matrices resulting in a block matrix and it is a generalization of the outer product.

## 2.2 Tensor Mixed Effects Model

Firstly, we consider a TME model for the third-order tensor data

$$\mathcal{Y}_i = \mathcal{F} \times_1 \boldsymbol{A}_i^{(1)} \times_2 \boldsymbol{A}_i^{(2)} \times_3 \boldsymbol{A}_i^{(3)} + \mathcal{R}_i \times_1 \boldsymbol{B}_i^{(1)} \times_2 \boldsymbol{B}_i^{(2)} \times_3 \boldsymbol{B}_i^{(3)} + \mathcal{E}_i \qquad (1)$$

where the $i^{\text{th}}$ response tensor is $\mathcal{Y}_i \in \mathbb{R}^{J \times K \times L}$ with $i = 1,\cdots,N$; $N$ is the sample size; the fixed effects core tensor is $\mathcal{F} \in \mathbb{R}^{P_1 \times Q_1 \times R_1}$; $\boldsymbol{A}_i^{(1)} \in \mathbb{R}^{J \times P_1}$, $\boldsymbol{A}_i^{(2)} \in \mathbb{R}^{K \times Q_1}$, $\boldsymbol{A}_i^{(3)} \in \mathbb{R}^{L \times R_1}$ are the design (factor) matrices for the fixed effects; the random effects core tensor is denoted by $\mathcal{R}_i \in \mathbb{R}^{P_2 \times Q_2 \times R_2}$, and the corresponding design (factor) matrices for the random effects by $\boldsymbol{B}_i^{(1)} \in \mathbb{R}^{J \times P_2}$, $\boldsymbol{B}_i^{(2)} \in \mathbb{R}^{K \times Q_2}$, $\boldsymbol{B}_i^{(3)} \in \mathbb{R}^{L \times R_2}$; the tensor for the residual errors is denoted by $\mathcal{E}_i \in \mathbb{R}^{J \times K \times L}$. Both the fixed effects and the random effects can be regarded as Tucker decompositions of original fixed/random effects. We also denote the Tucker decomposition by $[\![\mathcal{F}; A_i^{(1)}, A_i^{(2)}, A_i^{(3)}]\!] = \mathcal{F} \times_1 A_i^{(1)} \times_2 A_i^{(2)} \times_3 A_i^{(3)}$. Same as the requirement in Tucker decomposition, both the design matrices $A_i^{(j)}, j = 1,2,3$ and $B_i^{(j)}, j = 1,2,3$ are chosen to be orthogonal. Usually, the higher order Tucker decomposition follows similar structure as the third-order decomposition. Thus, it is straightforward to extend the third-order TME model to higher dimensional analysis.

Similar to the classical mixed effects model, we assume that the specification of the random effects core tensor $\mathcal{R}_i$ and the residual errors tensor $\mathcal{E}_i$ follow tensor normal distributions. Particularly, the tensor normal distribution of random effects core tensor $\mathcal{R}_i$ is $N_{P_2,Q_2,R_2}(\mathcal{O}; \Sigma_r, \Psi_r, \Omega_r)$, where the mean tensor $\mathcal{O}$ is a zero tensor, and the covariance matrices along different dimensions $\Sigma_r \in \mathbb{R}^{P_2 \times P_2}, \Psi_r \in \mathbb{R}^{Q_2 \times Q_2}, \Omega_r \in \mathbb{R}^{R_2 \times R_2}$ are positive definite. We know, from the properties of tensor normal distribution, $\text{vec}(\mathcal{R}_i)$ is distributed as a multivariate normal distribution with mean $\text{vec}(\mathcal{O})$ and covariance matrix $\Omega_r \otimes \Psi_r \otimes \Sigma_r$. We write that

$$\mathcal{R}_i \sim N_{P_2,Q_2,R_2}(\mathcal{O}; \Sigma_r, \Psi_r, \Omega_r) \text{ if } \text{vec}(\mathcal{R}_i) \sim N_{P_2 Q_2 R_2}(\text{vec}(\mathcal{O}), \Omega_r \otimes \Psi_r \otimes \Sigma_r).$$

Similarly, the distribution of the residual errors tensor $\mathcal{E}_i$ is $N_{J,K,L}(\mathcal{O}; \Sigma_\varepsilon, \Psi_\varepsilon, \Omega_\varepsilon)$, and the noise covariance matrices along different dimensions are $\Sigma_\varepsilon \in \mathbb{R}^{J \times J}, \Psi_\varepsilon \in \mathbb{R}^{K \times K}, \Omega_\varepsilon \in \mathbb{R}^{L \times L}$. Thus, $\text{vec}(\mathcal{E}_i) \sim N_{JKL}(\text{vec}(\mathcal{O}), \Omega_\varepsilon \otimes \Psi_\varepsilon \otimes \Sigma_\varepsilon)$. Moreover, we assume that the random effects core tensor and residual errors tensor are independent of each other. According to the descriptions above, we can find that the parameter size of the TME model in Equation (1) is $P_1 \times Q_1 \times R_1 + (P_2 + P_2^2 + Q_2 + Q_2^2 + R_2 + R_2^2)/2 + J + K + L$, with the assumption that the covariance matrices $\Sigma_\varepsilon, \Psi_\varepsilon, \Omega_\varepsilon$ are diagonal. While the parameter size of the corresponding vectorized linear mixed effects (vLME) model is $P_1 \times Q_1 \times R_1 + (P_2 Q_2 R_2 + P_2^2 Q_2^2 R_2^2)/2 + JKL$, with the assumption that the covariance matrix of noise term is diagonal. Therefore, the parameter size of the vLME model is much larger than the parameter size of the TME model.

In addition to the fixed effects core tensor and design matrices, the TME model includes two sources of random components: the random effects accounting for covariance along different dimensions, and the residual errors $\mathcal{E}_i$ relevant to the inevitable random noise. Based on the properties of tensor normal distribution, we can derive the random distribution of $\mathcal{Y}_i$, as shown in Proposition 1.

**Proposition 1**. The response tensor (1) follows a tensor normal distribution, that is

$$\mathcal{Y}_i \sim N_{J,K,L}\left(\llbracket \mathcal{F}; A_i^{(1)}, A_i^{(2)}, A_i^{(3)} \rrbracket; B_i^{(1)}\Sigma_r B_i^{(1)^T} + \Sigma_\varepsilon, B_i^{(2)}\Psi_r B_i^{(2)^T} + \Psi_\varepsilon, B_i^{(3)}\Omega_r B_i^{(3)^T} \right. \\ \left. + \Omega_\varepsilon \right) \quad (2)$$

Proof: please see the appendix A.1 in the supplementary materials.

For simplification, we define $\widetilde{\mathcal{F}} = \llbracket \mathcal{F}; A_i^{(1)}, A_i^{(2)}, A_i^{(3)} \rrbracket$, $\Sigma_i = B_i^{(1)}\Sigma_r B_i^{(1)^T} + \Sigma_\varepsilon$, $\Psi_i = B_i^{(2)}\Psi_r B_i^{(2)^T} + \Psi_\varepsilon$, and $\Omega_i = B_i^{(3)}\Omega_r B_i^{(3)^T} + \Omega_\varepsilon$. The total covariance matrices $\Sigma_i$, $\Psi_i$, $\Omega_i$ are positive definite. Thus, the response tensor distribution (2) can be written as $\mathcal{Y}_i \sim N_{J,K,L}(\widetilde{\mathcal{F}}; \Sigma_i, \Psi_i, \Omega_i)$. It can be further described in matrix form using three different modes as follows

$$Y_{i(1)} \sim N_{J,KL}\left(A_i^{(1)} F_{(1)}(A_i^{(3)} \otimes A_i^{(2)})^T; \Sigma_i, \Omega_i \otimes \Psi_i\right) \quad (3)$$

$$Y_{i(2)} \sim N_{K,JL}\left(A_i^{(2)} F_{(2)}(A_i^{(3)} \otimes A_i^{(1)})^T; \Psi_i, \Omega_i \otimes \Sigma_i\right) \quad (4)$$

$$Y_{i(3)} \sim N_{L,JK}\left(A_i^{(3)} F_{(3)}(A_i^{(2)} \otimes A_i^{(1)})^T; \Omega_i, \Psi_i \otimes \Sigma_i\right) \quad (5)$$

where $Y_{i(k)}$ and $F_{(k)}$ ($k$=1,2,3.) are the $k$-mode matricization of the tensor $\mathcal{Y}_i$ and $\mathcal{F}$. Obviously, Equations (3-5) show that $Y_{i(1)}$, $Y_{i(2)}$ and $Y_{i(3)}$ follow matrix normal distributions.

In this section, we proposed the TME model and specified the distribution of random effects core tensor and errors tensor. We also derived the random distribution of the response tensors in Proposition 1, which lays a foundation for inference in Section 3.

## 3. Inference of the TME Model

This section discusses how to estimate the parameters in the TME model by using the maximum likelihood estimation (MLE). Generally speaking, the parameter estimation of a TME model involves three steps: (i) constructing a log-likelihood function for the MLE with relevant probability distribution functions; (ii) deriving the MLE of fixed effects and total covariance matrices along different dimensions; and (iii) obtaining the MLE for covariance matrices of residual errors based on the conditional probability distribution.

## 3.1 Maximum Likelihood Estimation of Fixed Effects and Total Covariance Matrices

We know that the response tensor $\mathcal{Y}_i$ follows the tensor normal distribution in Equation (2) and the $k$-mode matricization $Y_{i(k)}$ follows the matrix normal distributions as shown in Equations (3-5). Thus, we have Proposition 2.

**Proposition 2**. The log-likelihood functions of Equations (3-5) are the same, and can be represented as

$$l_i = -\frac{JKL}{2}\log 2\pi - \frac{JK}{2}\log|\Omega_i| - \frac{JL}{2}\log|\Psi_i| - \frac{KL}{2}\log|\Sigma_i| - \frac{1}{2}\left(\text{vec}\left(Y_{i(1)} - A_i^{(1)}F_{(1)}\left(A_i^{(3)} \otimes A_i^{(2)}\right)^T\right)\right)^T (\Omega_i^{-1} \otimes \Psi_i^{-1} \otimes \Sigma_i^{-1})\text{vec}\left(Y_{i(1)} - A_i^{(1)}F_{(1)}\left(A_i^{(3)} \otimes A_i^{(2)}\right)^T\right). \quad (6)$$

Proof: please see the appendix A.2 in the supplementary materials.

Maximization of the log-likelihood function (6) yields the MLE estimators, which are shown in Proposition 3.

**Proposition 3**. Given the response tensors $\mathcal{Y}_i$ and the basis $A_i^{(1)}, A_i^{(2)}, A_i^{(3)}$ with $i = 1, \cdots, N$, the maximum likelihood estimator of $\text{vec}(\mathcal{F})$ is

$$\text{vec}(\widehat{\mathcal{F}}) = \left(\sum_{i=1}^{N}\left(A_i^{(3)^T}\Omega_i^{-1}A_i^{(3)}\right) \otimes \left(A_i^{(2)^T}\Psi_i^{-1}A_i^{(2)}\right) \otimes \left(A_i^{(1)^T}\Sigma_i^{-1}A_i^{(1)}\right)\right)^{-1} \cdot \left(\sum_{i=1}^{N}\left(A_i^{(3)^T}\Omega_i^{-1}\right) \otimes \left(A_i^{(2)^T}\Psi_i^{-1}\right) \otimes \left(A_i^{(1)^T}\Sigma_i^{-1}\right) \cdot \text{vec}(\mathcal{Y}_i)\right) \quad (7)$$

When $B_i^{(1)}, B_i^{(2)},$ and $B_i^{(3)}$ are constant for all $i = 1, \cdots, N$, and setting $B_i^{(1)} = B^{(1)}, B_i^{(2)} = B^{(2)}, B_i^{(3)} = B^{(3)}$ for $i = 1, \cdots, N$. For simplification, we define $\widehat{\widehat{\mathcal{F}}} = [\![\widehat{\mathcal{F}}; A_i^{(1)}, A_i^{(2)}, A_i^{(3)}]\!]$. The maximum likelihood estimators of $\Sigma_i, \Psi_i, \Omega_i$ are

$$\widehat{\Sigma}_i = \frac{1}{KLN}\sum_{i=1}^{N}\left(\mathcal{Y}_i - \widehat{\widehat{\mathcal{F}}}\right)_{(1)} \cdot \left(\widehat{\Omega}_i^{-1} \otimes \widehat{\Psi}_i^{-1}\right) \cdot \left(\mathcal{Y}_i - \widehat{\widehat{\mathcal{F}}}\right)_{(1)}^T \quad (8)$$

$$\widehat{\Psi}_i = \frac{1}{JLN}\sum_{i=1}^{N}\left(\mathcal{Y}_i - \widehat{\widehat{\mathcal{F}}}\right)_{(2)} \cdot \left(\widehat{\Omega}_i^{-1} \otimes \widehat{\Sigma}_i^{-1}\right) \cdot \left(\mathcal{Y}_i - \widehat{\widehat{\mathcal{F}}}\right)_{(2)}^T \quad (9)$$

$$\widehat{\boldsymbol{\Omega}}_i = \frac{1}{JKN}\sum_{i=1}^{N}\left(\boldsymbol{\mathcal{Y}}_i - \widehat{\widetilde{\boldsymbol{\mathcal{F}}}}\right)_{(3)} \cdot \left(\widehat{\boldsymbol{\Psi}}_i^{-1} \otimes \widehat{\boldsymbol{\Sigma}}_i^{-1}\right) \cdot \left(\boldsymbol{\mathcal{Y}}_i - \widehat{\widetilde{\boldsymbol{\mathcal{F}}}}\right)_{(3)}^{T} \quad (10)$$

If both $(\boldsymbol{A}_i^{(1)}, \boldsymbol{A}_i^{(2)}, \boldsymbol{A}_i^{(3)})$ and $(\boldsymbol{B}_i^{(1)}, \boldsymbol{B}_i^{(2)}, \boldsymbol{B}_i^{(3)})$ are constant for all $i = 1, \cdots, N$, $\overline{\boldsymbol{\mathcal{Y}}}$ is the mean response tensor, the maximum likelihood estimators of $\boldsymbol{\Sigma}_i, \boldsymbol{\Psi}_i, \boldsymbol{\Omega}_i$ are

$$\widehat{\boldsymbol{\Sigma}}_i = \frac{1}{KLN}\sum_{i=1}^{N}(\boldsymbol{\mathcal{Y}}_i - \overline{\boldsymbol{\mathcal{Y}}})_{(1)} \cdot \left(\widehat{\boldsymbol{\Omega}}_i^{-1} \otimes \widehat{\boldsymbol{\Psi}}_i^{-1}\right) \cdot (\boldsymbol{\mathcal{Y}}_i - \overline{\boldsymbol{\mathcal{Y}}})_{(1)}^{T} \quad (11)$$

$$\widehat{\boldsymbol{\Psi}}_i = \frac{1}{JLN}\sum_{i=1}^{N}(\boldsymbol{\mathcal{Y}}_i - \overline{\boldsymbol{\mathcal{Y}}})_{(2)} \cdot \left(\widehat{\boldsymbol{\Omega}}_i^{-1} \otimes \widehat{\boldsymbol{\Sigma}}_i^{-1}\right) \cdot (\boldsymbol{\mathcal{Y}}_i - \overline{\boldsymbol{\mathcal{Y}}})_{(2)}^{T} \quad (12)$$

$$\widehat{\boldsymbol{\Omega}}_i = \frac{1}{JKN}\sum_{i=1}^{N}(\boldsymbol{\mathcal{Y}}_i - \overline{\boldsymbol{\mathcal{Y}}})_{(3)} \cdot \left(\widehat{\boldsymbol{\Psi}}_i^{-1} \otimes \widehat{\boldsymbol{\Sigma}}_i^{-1}\right) \cdot (\boldsymbol{\mathcal{Y}}_i - \overline{\boldsymbol{\mathcal{Y}}})_{(3)}^{T} \quad (13)$$

Proof: please see the appendix A.3 in the supplementary materials.

Moreover, we can show that the estimator $\text{vec}(\widehat{\boldsymbol{\mathcal{F}}})$ given in Equation (7) is uniquely determined regardless of the parametrization of the covariance matrices, which is explored in appendix A.3.

From Equations (8-13), we can see that the estimators of covariance matrices are cross-related. A Flip-Flop type algorithm is designed to compute them. We will describe the algorithm in Section 4. Before that, we continue to explore the MLE for the covariance matrices of residual errors based on the conditional probability distributions.

### 3.2 Maximum Likelihood Estimation of Random Effects and Residual Covariance Matrices

After finishing the estimation of the fixed effects and total covariance matrices, we consider the estimation for the covariance matrices of random effects and residual errors. The distribution of random effects core tensor $\boldsymbol{\mathcal{R}}_i$ conditional on response tensors $\boldsymbol{\mathcal{Y}}_i$ ($i = 1, \cdots, N$) follows a tensor normal distribution. Assuming $\boldsymbol{\mathcal{Y}}_i$ and $\widetilde{\boldsymbol{\mathcal{F}}}$ are known, the estimation of $\boldsymbol{\mathcal{R}}_i$ is the expectation of $\boldsymbol{\mathcal{R}}_i|\boldsymbol{\mathcal{Y}}_i$ and it can obtain

$$\widehat{\boldsymbol{\mathcal{R}}}_i = \left[\!\left[\boldsymbol{\mathcal{Y}}_i - \widetilde{\boldsymbol{\mathcal{F}}}; \boldsymbol{\Sigma}_{\text{r}}\boldsymbol{B}_i^{(1)^T}\boldsymbol{\Sigma}_i^{-1}, \boldsymbol{\Psi}_{\text{r}}\boldsymbol{B}_i^{(2)^T}\boldsymbol{\Psi}_i^{-1}, \boldsymbol{\Omega}_{\text{r}}\boldsymbol{B}_i^{(3)^T}\boldsymbol{\Omega}_i^{-1}\right]\!\right]. \quad (14)$$

The distribution of $\boldsymbol{\mathcal{Y}}_i - \widetilde{\boldsymbol{\mathcal{F}}} - [\![\boldsymbol{\mathcal{R}}_i; \boldsymbol{B}_i^{(1)}, \boldsymbol{B}_i^{(2)}, \boldsymbol{B}_i^{(3)}]\!]$ conditioned on the random effects core tensor $\boldsymbol{\mathcal{R}}_i$ is a tensor normal distribution given by

$$\left(\boldsymbol{\mathcal{Y}}_i - \widetilde{\boldsymbol{\mathcal{F}}} - [\![\boldsymbol{\mathcal{R}}_i; \boldsymbol{B}_i^{(1)}, \boldsymbol{B}_i^{(2)}, \boldsymbol{B}_i^{(3)}]\!]\right) | \boldsymbol{\mathcal{R}}_i \sim N_{J,K,L}(\boldsymbol{\mathcal{O}}; \boldsymbol{\Sigma}_\varepsilon, \boldsymbol{\Psi}_\varepsilon, \boldsymbol{\Omega}_\varepsilon).$$

For simplification, we define $\widetilde{\boldsymbol{\mathcal{R}}}_i = [\![\boldsymbol{\mathcal{R}}_i; \boldsymbol{B}_i^{(1)}, \boldsymbol{B}_i^{(2)}, \boldsymbol{B}_i^{(3)}]\!]$, $\widehat{\widetilde{\boldsymbol{\mathcal{R}}}}_i = [\![\widehat{\boldsymbol{\mathcal{R}}}_i; \boldsymbol{B}_i^{(1)}, \boldsymbol{B}_i^{(2)}, \boldsymbol{B}_i^{(3)}]\!]$. Similar to Proposition 3, we have the maximum likelihood estimators of $\boldsymbol{\Sigma}_\varepsilon, \boldsymbol{\Psi}_\varepsilon, \boldsymbol{\Omega}_\varepsilon$ are

$$\widehat{\boldsymbol{\Sigma}}_\varepsilon = \frac{1}{KLN} \sum_{i=1}^{N} \left(\boldsymbol{\mathcal{Y}}_i - \widehat{\widetilde{\boldsymbol{\mathcal{F}}}} - \widehat{\widetilde{\boldsymbol{\mathcal{R}}}}\right)_{(1)} \cdot (\widehat{\boldsymbol{\Omega}}_\varepsilon^{-1} \otimes \widehat{\boldsymbol{\Psi}}_\varepsilon^{-1}) \cdot \left(\boldsymbol{\mathcal{Y}}_i - \widehat{\widetilde{\boldsymbol{\mathcal{F}}}} - \widehat{\widetilde{\boldsymbol{\mathcal{R}}}}\right)_{(1)}^T \tag{15}$$

$$\widehat{\boldsymbol{\Psi}}_\varepsilon = \frac{1}{JLN} \sum_{i=1}^{N} \left(\boldsymbol{\mathcal{Y}}_i - \widehat{\widetilde{\boldsymbol{\mathcal{F}}}} - \widehat{\widetilde{\boldsymbol{\mathcal{R}}}}\right)_{(2)} \cdot (\widehat{\boldsymbol{\Omega}}_\varepsilon^{-1} \otimes \widehat{\boldsymbol{\Sigma}}_\varepsilon^{-1}) \cdot \left(\boldsymbol{\mathcal{Y}}_i - \widehat{\widetilde{\boldsymbol{\mathcal{F}}}} - \widehat{\widetilde{\boldsymbol{\mathcal{R}}}}\right)_{(2)}^T \tag{16}$$

$$\widehat{\boldsymbol{\Omega}}_\varepsilon = \frac{1}{JKN} \sum_{i=1}^{N} \left(\boldsymbol{\mathcal{Y}}_i - \widehat{\widetilde{\boldsymbol{\mathcal{F}}}} - \widehat{\widetilde{\boldsymbol{\mathcal{R}}}}\right)_{(3)} \cdot (\widehat{\boldsymbol{\Psi}}_\varepsilon^{-1} \otimes \widehat{\boldsymbol{\Sigma}}_\varepsilon^{-1}) \cdot \left(\boldsymbol{\mathcal{Y}}_i - \widehat{\widetilde{\boldsymbol{\mathcal{F}}}} - \widehat{\widetilde{\boldsymbol{\mathcal{R}}}}\right)_{(3)}^T \tag{17}$$

Comparing Equations (15-17) with Equations (8-10), we notice similar patterns. The mean components change from $\boldsymbol{\mathcal{Y}}_i - \widehat{\widetilde{\boldsymbol{\mathcal{F}}}}$ to $\boldsymbol{\mathcal{Y}}_i - \widehat{\widetilde{\boldsymbol{\mathcal{F}}}} - \widehat{\widetilde{\boldsymbol{\mathcal{R}}}}$ based on the estimation of random effects. After that, a progressive estimation for covariance matrices $\boldsymbol{\Sigma}_\varepsilon, \boldsymbol{\Psi}_\varepsilon$ and $\boldsymbol{\Omega}_\varepsilon$ are obtained.

We know that the covariance matrices $\boldsymbol{\Sigma}_i, \boldsymbol{\Psi}_i, \boldsymbol{\Omega}_i$ and $\boldsymbol{\Sigma}_\varepsilon, \boldsymbol{\Psi}_\varepsilon, \boldsymbol{\Omega}_\varepsilon$ should be positive definite. In order to ensure the positive definite property in Equations (8-10, 15-17), the existence of the MLE should be explored. This is shown in Section 3.2. Based on $\boldsymbol{\Sigma}_i = \boldsymbol{B}_i^{(1)} \boldsymbol{\Sigma}_r \boldsymbol{B}_i^{(1)^T} + \boldsymbol{\Sigma}_\varepsilon$, $\boldsymbol{\Psi}_i = \boldsymbol{B}_i^{(2)} \boldsymbol{\Psi}_r \boldsymbol{B}_i^{(2)^T} + \boldsymbol{\Psi}_\varepsilon$, $\boldsymbol{\Omega}_i = \boldsymbol{B}_i^{(3)} \boldsymbol{\Omega}_r \boldsymbol{B}_i^{(3)^T} + \boldsymbol{\Omega}_\varepsilon$, we know that the covariance matrices of random effects and residual errors are not unique. The identifiability should be investigated, which is discussed in Section 3.3.

### 3.2 Existence of the MLE

Finding the estimation of the average component, fixed effects core tensor $\widehat{\boldsymbol{\mathcal{F}}}$, is straightforward given the positive definite covariance matrices. Hereafter, we focus on the exploration of the existence of MLE for the total covariance matrices $\widehat{\boldsymbol{\Sigma}}_i, \widehat{\boldsymbol{\Psi}}_i, \widehat{\boldsymbol{\Omega}}_i$, shown in Equations (8-13). A necessary condition for the existence of the MLE can be derived based on the paper (Manceur and Dutilleul 2013), which is demonstrated in Proposition 4.

**Proposition 4**. If maximum likelihood estimators for the covariance matrices $\boldsymbol{\Sigma}_i, \boldsymbol{\Psi}_i, \boldsymbol{\Omega}_i$ in the TME model (1) exist, the sample size $N$ of the response tensors $\boldsymbol{\mathcal{Y}}_i$ $(i = 1, \cdots, N)$ satisfies the condition

$$N \geq \max\left(\frac{J}{KL}, \frac{K}{JL}, \frac{L}{JK}\right) + 1.$$

The Proof is straightforward according to the conclusion in the paper (Manceur and Dutilleul 2013). Although the condition shown in Proposition 4 is necessary for the existence of the MLE, it is not sufficient because it cannot ensure that all the iterations of the algorithm have full rank matrices. Similar to the existence of the MLE for the model with Kronecker product covariance structure (Roś et al. 2016), it could happen that covariance matrices in the updated iterations do not have a full rank with the likelihood of the TME model converging to the supremum. The reason is that the space $\{\boldsymbol{\Omega}_i \otimes \boldsymbol{\Psi}_i \otimes \boldsymbol{\Sigma}_i : \boldsymbol{\Sigma}_i \in \mathbb{R}^{J \times J}, \boldsymbol{\Psi}_i \in \mathbb{R}^{K \times K}, \boldsymbol{\Omega}_i \in \mathbb{R}^{L \times L}; \boldsymbol{\Omega}_i, \boldsymbol{\Psi}_i, \boldsymbol{\Sigma}_i$ are positive definite$\}$ with any norm is not closed. If we choose a stronger condition, for a space $\mathbb{K}$ (equipped with the Frobenius norm) of positive definite $JKL \times JKL$ matrices that have a kronecker structure such that $\boldsymbol{\Omega}_i \otimes \boldsymbol{\Psi}_i \otimes \boldsymbol{\Sigma}_i \in \mathbb{K}$, where $\boldsymbol{\Omega}_i, \boldsymbol{\Psi}_i, \boldsymbol{\Sigma}_i$ are also positive definite, Then $\mathbb{K}$ is closed, according to the natural extension of (Roś et al. 2016). Based on this conclusion, we can formulate the sufficient condition for the existence of the MLE for the total covariance matrices, as shown in Proposition 5.

**Proposition 5**. The response tensor $\boldsymbol{\mathcal{Y}}_i$ ($i = 1, \cdots, N$) satisfies the model (1). If $N \geq JKL$, maximum likelihood estimators for the covariance matrices $\boldsymbol{\Sigma}_i$, $\boldsymbol{\Psi}_i$, $\boldsymbol{\Omega}_i$ in the TME model exist with probability 1.

The proof is straightforward by using the conclusion (Burg et al. 1982), and it is an extension of Theorem 3 in page 6 of Roś et al. (2016).

In summary of Propositions 4 and 5, if $N < \max\left(\frac{J}{KL}, \frac{K}{JL}, \frac{L}{JK}\right) + 1$, the MLEs of covariance matrices do not exist. However, if $N \geq JKL$, the MLEs exist with probability 1.

Moreover, the dimensions of tensor samples are usually large, and $N \geq JKL$ is hard to guarantee in practice. When the covariance matrices satisfy special structures, such as diagonal structures, the sufficient condition of existence will be changed. We have Proposition 6 to illustrate the existence conditions with the additional assumptions of diagonal matrices.

**Proposition 6**. The response tensors $\boldsymbol{\mathcal{Y}}_i$ ($i = 1, \cdots, N$) satisfy the model (1) with the additional assumption that $\boldsymbol{\Sigma}_i$ is diagonal. If $N \geq \max\left(KL, \max\left(\frac{J}{KL}, \frac{K}{JL}, \frac{L}{JK}\right) + 1\right)$, the maximum likelihood estimators for the covariance matrices $\boldsymbol{\Sigma}_i, \boldsymbol{\Psi}_i, \boldsymbol{\Omega}_i$ in the TME model exist with probability 1.

Proposition 6 is a three-dimensional extension of Theorem 8 on page 14 of Roś et al. (2016).

Due to the similarity of the MLE for covariance matrices $\boldsymbol{\Sigma}_i, \boldsymbol{\Psi}_i, \boldsymbol{\Omega}_i$ and $\boldsymbol{\Sigma}_\varepsilon, \boldsymbol{\Psi}_\varepsilon, \boldsymbol{\Omega}_\varepsilon$, the existence condition of the MLE for covariance matrices $\boldsymbol{\Sigma}_\varepsilon, \boldsymbol{\Psi}_\varepsilon, \boldsymbol{\Omega}_\varepsilon$ can be obtained accordingly. We need to point out that the existence of MLE does not mean that the estimation has good identifiability and convergence to the global optimal solution. We will investigate the identifiability in Section 3.3 and convergence in Section 4.2.

### 3.3 Identifiability

The identifiability of a statistical model is essential because it ensures correct inference on model parameters. For the TME model, the identifiability is extremely complex because it involves three aspects: (i) whether the fixed effects core tensor is identifiable; (ii) the identifiability of the Kronecker covariance structure, because $\boldsymbol{\Omega}_i \otimes \boldsymbol{\Psi}_i = c\boldsymbol{\Omega}_i \otimes \frac{1}{c}\boldsymbol{\Psi}_i$ for any $c > 0$; (iii) and the identifiability of covariance matrices of random effects and residual errors, because $\boldsymbol{\Sigma}_i = \boldsymbol{B}_i^{(1)} \boldsymbol{\Sigma}_r \boldsymbol{B}_i^{(1)^T} + \boldsymbol{\Sigma}_\varepsilon$, $\boldsymbol{\Psi}_i = \boldsymbol{B}_i^{(2)} \boldsymbol{\Psi}_r \boldsymbol{B}_i^{(2)^T} + \boldsymbol{\Psi}_\varepsilon$ and $\boldsymbol{\Omega}_i = \boldsymbol{B}_i^{(3)} \boldsymbol{\Omega}_r \boldsymbol{B}_i^{(3)^T} + \boldsymbol{\Omega}_\varepsilon$. We will investigate the identifiability for each these aspects respectively.

Firstly, the identifiability of the TME model follows the identifiability definition of a linear mixed effects model (Demidenko 2013). If the TME model is defined by a family of distributions $\{P_{\boldsymbol{\theta}}, \boldsymbol{\theta} \in \boldsymbol{\Theta}\}$, as shown in Equation (2), which is parameterized by the vector $\boldsymbol{\theta}$, and $\boldsymbol{\Theta}$ is the parameter space. The model is identifiable on $\boldsymbol{\Theta}$ if $P_{\boldsymbol{\theta}_1} = P_{\boldsymbol{\theta}_2}$ implies that $\boldsymbol{\theta}_1 = \boldsymbol{\theta}_2$. Identifiability is a necessary property for the adequacy of the TME model.

In the linear mixed effects model, the design matrix for fixed effects has to be full-ranked to realize unique estimation of fixed effects parameters. If $[\![\boldsymbol{\mathcal{F}}_1; \boldsymbol{A}_i^{(1)}, \boldsymbol{A}_i^{(2)}, \boldsymbol{A}_i^{(3)}]\!] = [\![\boldsymbol{\mathcal{F}}_2; \boldsymbol{A}_i^{(1)}, \boldsymbol{A}_i^{(2)}, \boldsymbol{A}_i^{(3)}]\!]$ implies $\boldsymbol{\mathcal{F}}_1 = \boldsymbol{\mathcal{F}}_2$, it means the fixed effects core tensor is indentifiable. To ensure the identifiability of fixed effects core tensor in the TME model, it must satisfy that the design matrices $\boldsymbol{A}_i^{(1)}, \boldsymbol{A}_i^{(2)}, \boldsymbol{A}_i^{(3)}$ have full rank. According to the property of Kronecker product, $\boldsymbol{A}_i^{(3)} \otimes \boldsymbol{A}_i^{(2)} \otimes \boldsymbol{A}_i^{(1)}$ has full rank.

The identifiability of the Kronecker covariance structure can be ensured by introducing additional constraints. Because the covariance matrices are positive definite, one kind of constraint can be fixing particular summations of the diagonal elements of $\boldsymbol{\Sigma}_i, \boldsymbol{\Psi}_i$ or $\boldsymbol{\Omega}_i$ to be equal to 1; another possible constraint is to assume that the determinants of two of the three covariance matrices are equal to 1. The constraints do not restrict the application of the TME model since the relative magnitude of the entries in the covariance matrix will be relevant to the key information that we care about.

Based on $\Sigma_i = B_i^{(1)} \Sigma_r B_i^{(1)T} + \Sigma_\varepsilon$, $\Psi_i = B_i^{(2)} \Psi_r B_i^{(2)T} + \Psi_\varepsilon$ and $\Omega_i = B_i^{(3)} \Omega_r B_i^{(3)T} + \Omega_\varepsilon$, we know that the covariance matrices of random effects and residual errors are not unique. In order to ensure identifiability in a similar way to the classical mixed effects model (Demidenko 2013), we need to ensure that the design matrices $B_i^{(1)}, B_i^{(2)}, B_i^{(3)}$ have full rank and to specify the structure of the covariance matrices $\Sigma_\varepsilon, \Psi_\varepsilon, \Omega_\varepsilon$. In general, there are two ways to specify the structure of the covariance matrices: One way is to assume the covariance matrices $\Sigma_\varepsilon, \Psi_\varepsilon, \Omega_\varepsilon$ are diagonal matrices (corresponding to the independent noise). Notably, the covariance matrices for random effects $\Sigma_r, \Psi_r, \Omega_r$ and the total covariance matrices $\Sigma_i, \Psi_i, \Omega_i$ are not diagonal. Another way is to determine the noise pattern based on a phase-I analysis. For example the noise is found to have a signal dependent property and the noise parameters are consistent for different data acquisition time in Raman inspection of nanomanufacturing (Yue et al. 2017).

## 4. Double Flip-Flop Algorithm for Parameter Estimation of TME Model
### 4.1 Double Flip-Flop Algorithm

For existing maximum likelihood estimation of covariance matrices with Kronecker structure, a Flip-Flop algorithm has been proposed to update the estimation of several components sequentially and iteratively (Dutilleul 1999; Lu and Zimmerman 2004; Manceur and Dutilleul 2013; Sakata 2016). We have derived the maximum likelihood estimators for the TME model in Section 3. In this section, we will propose a double Flip-Flop algorithm to implement the parameter estimation iteratively. The algorithm is shown in Table 1.

**Table 1: Double Flip-Flop Algorithm for the TME Model**

Step 1: Initialize the core tensor $\widehat{\mathcal{F}}^{\{0\}}$ and design matrices $A_i^{(1)}, A_i^{(2)}, A_i^{(3)}, B_i^{(1)}, B_i^{(2)}, B_i^{(3)}$, and covariance matrices $\widehat{\Sigma}_r^{\{0\}}, \widehat{\Psi}_r^{\{0\}}, \widehat{\Omega}_r^{\{0\}}, \widehat{\Sigma}_\varepsilon^{\{0\}}, \widehat{\Psi}_\varepsilon^{\{0\}}, \widehat{\Omega}_\varepsilon^{\{0\}}$. Set the iteration number $k = 0$.

➢ Calculate the mean response tensor $\overline{\mathcal{Y}}$, and then use high-order orthogonal iteration (HOOI) to compute a rank-$(P_1, Q_1, R_1)$ Tucker decomposition, $\overline{\mathcal{Y}} = [\![\mathcal{F}^{\{0\}}; A_i^{(1)\{0\}}, A_i^{(2)\{0\}}, A_i^{(3)\{0\}}]\!]$. The decomposed core tensor and factor matrices work as the initialized fixed effect core tensor and design matrices for fixed effects.

| |
|---|
| ➢ Choose the design matrices for random effects $B_i^{(1)\{0\}}, B_i^{(2)\{0\}}, B_i^{(3)\{0\}}$ as a subset of appropriate columns of the design matrices $A_i^{(1)\{0\}}, A_i^{(2)\{0\}}, A_i^{(3)\{0\}}$. <br> ➢ Compute the $\widehat{\Sigma}_i^{\{0\}}, \widehat{\Psi}_i^{\{0\}}, \widehat{\Omega}_i^{\{0\}}$. |
| Step 2: Increase iteration number $k$ by 1. |
| Step 3: Keep $\widehat{\mathcal{F}}^{\{k-1\}}$ fixed and compute $\widehat{\Sigma}_i^{\{k\}}, \widehat{\Psi}_i^{\{k\}}, \widehat{\Omega}_i^{\{k\}}$. <br> ➢ Compute $\widehat{\Sigma}_i^{\{k\}}$ by using Equation (8) (using $\Psi_i^{\{k-1\}}, \widehat{\Omega}_i^{\{k-1\}}$). <br> ➢ Compute $\widehat{\Psi}_i^{\{k\}}$ by using Equation (9) (using $\widehat{\Sigma}_i^{\{k\}}, \widehat{\Omega}_i^{\{k-1\}}$). <br> ➢ Compute $\widehat{\Omega}_i^{\{k\}}$ by using Equation (10) (using $\widehat{\Sigma}_i^{\{k\}}, \widehat{\Psi}_i^{\{k\}}$). |
| Step 4: Keep $\widehat{\Sigma}_i^{\{k\}}, \widehat{\Psi}_i^{\{k\}}, \widehat{\Omega}_i^{\{k\}}$ fixed and compute $\widehat{\mathcal{F}}^{\{k\}}$ by using Equation (7). |
| Step 5: Iterate between steps 2 and 4 until convergence or until reaching a predetermined number of iterations $K$. |
| Step 6: Set the iteration number $t = 0$. Estimate $\widehat{\mathcal{R}}^{\{0\}}$ by the expectation mean in Equation (14). |
| Step 7: Increase iteration number $t$ by 1. |
| Step 8: Keep $\widehat{\mathcal{F}}^{\{k\}}, \widehat{\mathcal{R}}^{\{t-1\}}$ fixed and compute $\widehat{\Sigma}_\varepsilon, \widehat{\Psi}_\varepsilon, \widehat{\Omega}_\varepsilon$ considering given constraints. <br> ➢ Compute $\widehat{\Sigma}_\varepsilon^{\{t\}}$ by using Equation (15) (using $\widehat{\Psi}_\varepsilon^{\{t-1\}}, \widehat{\Omega}_\varepsilon^{\{t-1\}}$) and adjust it according to given constraints. <br> ➢ Compute $\widehat{\Psi}_\varepsilon^{\{t\}}$ by using Equation (16) (using $\widehat{\Sigma}_\varepsilon^{\{t\}}, \widehat{\Omega}_\varepsilon^{\{t-1\}}$) and adjust it according to given constraints. <br> ➢ Compute $\widehat{\Omega}_\varepsilon^{\{t\}}$ by using Equation (17) (using $\widehat{\Sigma}_\varepsilon^{\{t\}}, \widehat{\Psi}_\varepsilon^{\{t\}}$) and adjust it according to given constraints. |
| Step 9: Keep $\widehat{\Sigma}_\varepsilon^{\{t\}}, \widehat{\Psi}_\varepsilon^{\{t\}}, \widehat{\Omega}_\varepsilon^{\{t\}}$ fixed and compute $\widehat{\mathcal{R}}^{\{t\}}$ by using Equation (14). |
| Step 10: Iterate between steps 7 and 9 until convergence or until reaching a predetermined number of iterations $T$. |

Note that the algorithm involves two iterative loops. The first one is related to the computation of the fixed effects and total covariance matrices, and the second one is relevant to the computation of covariance matrices of residual errors and random effects. Each loop follows the characteristic of a Flip-Flop algorithm, and that is why it is named after the double Flip-Flop algorithm.

### 4.2 Initialization of the Algorithm

Obtaining good initial values is important for parameter estimation in the TME model. For the initialization, we use high-order orthogonal iteration (HOOI) to compute a rank-$(P_1, Q_1, R_1)$ Tucker decomposition. In the HOOI algorithm, a higher-order Singular Value Decomposition (HOSVD) is applied to initialize the factor matrices, and a set of orthogonal constraints to ensure the core tensor is all-orthogonal. This improves the uniqueness of Tucker decomposition (Kolda and Bader 2009). It is typically a challenging task to determine the parameters $P_1, Q_1, R_1$. Basically, the parameters $P_1, Q_1, R_1$ should be relevant to the rank of given tensor. However, there is no straightforward algorithm to determine the rank of a specific given tensor, which is NP-hard (Hastad 1990, Kolda and Bader 2009). In the implementations, we determine the key dimensional parameters by using the following procedures: (1) with the assumption that only part of features have random effects, we test that the ranges of these parameters $P_1, Q_1, R_1$ are J: K: L~1:1:1. The ranges of parameters $P_2, Q_2, R_2$ are $P_1: Q_1: R_1$~1:1:1. (2) we conduct tensor decomposition for each combination of parameters. (3) after obtaining core tensor from tensor decomposition, we check the sparsity for each combination of parameters. The sparsity is usually represented by the number of far-from-zero entries. For example, we choose a sparsity criterion that the summation of absolute value at one row of core tensor should be larger than a specific threshold. (4) we select the key dimensional parameters that have the largest summation of these parameters, as well as satisfy the sparsity criterion. The design matrices $A_i^{(1)\{0\}}, A_i^{(2)\{0\}}, A_i^{(3)\{0\}}$ are determined by the factor matrices from the Tucker decomposition. The design matrices for random effects $B_i^{(1)\{0\}}, B_i^{(2)\{0\}}, B_i^{(3)\{0\}}$ can be chosen as a subset of appropriate columns of the design matrices $A_i^{(1)\{0\}}, A_i^{(2)\{0\}}, A_i^{(3)\{0\}}$. While the columns are determined by possible random effects relevant to features of interest in a phase-I data analysis. In phase-I data analysis, a set of process data is gathered and analyzed all at once in a retrospective analysis, and the features of interest will be chosen by multiple trials.

**4.3 Convergence of the Algorithm**

Lu and Zimmerman (2004) have explored the convergence of a Flip-Flop algorithm. According to the paper (Lu and Zimmerman 2004), the likelihood function of successive iterations of a Flip-Flop algorithm cannot decrease. Provided $N \geq JKL$, the algorithm is guaranteed to converge. However, whether it converges to a MLE is not ensured because the space of the covariance matrices is not convex. An empirical study of the convergence is investigated in Section 5.1.

The most commonly used stopping criteria are ones that based on the relative change in either the covariance parameters between successive iterations or differences between successive log-likelihood functions. Considering all the covariance matrices, the stopping criteria for the first loop are that the $L_1$ norms $\left\|\widehat{\Sigma}_i^{\{k\}} - \widehat{\Sigma}_i^{\{k-1\}}\right\|_1, \left\|\widehat{\Psi}_i^{\{k\}} - \widehat{\Psi}_i^{\{k-1\}}\right\|_1, \left\|\widehat{\Omega}_i^{\{k\}} - \widehat{\Omega}_i^{\{k-1\}}\right\|_1$ are simultaneously smaller than the thresholds. Similar stopping criteria are applied for the second

loop, which means that $\left\|\widehat{\boldsymbol{\Sigma}}_{\varepsilon}^{\{t\}} - \widehat{\boldsymbol{\Sigma}}_{\varepsilon}^{\{t-1\}}\right\|_1$, $\left\|\widehat{\boldsymbol{\Psi}}_{\varepsilon}^{\{t\}} - \widehat{\boldsymbol{\Psi}}_{\varepsilon}^{\{t-1\}}\right\|_1$, $\left\|\widehat{\boldsymbol{\Omega}}_{\varepsilon}^{\{t\}} - \widehat{\boldsymbol{\Omega}}_{\varepsilon}^{\{t-1\}}\right\|_1$ are simultaneously smaller than the thresholds. For the asymptotic properties of the Flip-Flop type algorithm, please refer to the paper (Werner et al. 2008). We also investigate the asymptotic properties in Sections 5 through simulation and surrogated data analysis.

### 4.4 Computational Complexity of the Algorithm

Since the double Flip-Flop algorithm uses the HOOI algorithm to do initialization and then conducts two iterative Flip-Flop loops, we need to analyze the computational cost of this algorithm. For simplicity, we assume the dimensions $J = K = L$, $P_1 = Q_1 = R_1$, $P_2 = Q_2 = R_2$. In the initialization part, each iteration in HOOI involves six tensor-by-matrix products and three maximization problems, where the computational complexity for each HOOI iteration is $O(J^3 P_1 + J P_1^4 + P_1^6)$ (Elden and Savas 2009). The computational complexity of step 3 is $O(NJ^5)$, while step 4 is $O(NP_1^9 + NJ^3 P_1^3)$. Therefore the cost for each iteration of the first Flip-Flop loop is $O(NJ^5 + NP_1^9 + NJ^3 P_1^3)$. Similarly, the computational complexities for step 8 and step 9 are $O(NJ^5)$ and $O(NJ^3 P_2)$, respectively. Thus, the computation cost for each iteration of the second Flip-Flop loop is $O(NJ^5)$, which is dominated by step 8. The computational time will also be impacted by the iteration number. According to the simulation study in Section 5.1, the algorithm will converge quickly.

To show the computational advantage of the proposed TME model, we consider the conventional linear mixed effects model for vectorized responses (marked as vLME) (Galecki and Burzykowski 2013). After vectorization of the tensor responses, the dimension of each response becomes $J^3$. Thus the computational complexity of the vLME model is $O(NJ^9)$ (Lippert et al. 2011). It is much larger than the complexity of the TME model, which is $O(NJ^5 + NP_1^9 + NJ^3 P_1^3)$.

### 5. Numerical Analysis

### 5.1 Simulation Study

In this section, the performance of the iterative algorithm is evaluated through simulation studies. In order to simulate the response tensor with mixed effects, we generate the fixed effects tensor with dimension 30×5×5. The dimensions of core tensors for fixed effects and random effects are 8×3×3 and 3×2×2, respectively. The covariance matrices of random effects are generated from random symmetric positive definite matrices. Two covariance matrices of residual errors are generated by isotropic matrices (an isotropic matrix is an identity matrix multiplied by a positive number) with dimension 5×5 and another covariance matrix with dimension 30×30. 1000 response tensors are generated to test the performance. A computer with Intel Core i7-4500U processor and 8.00GB RAM is used to conduct the numerical analysis.

After we generate the dataset, we run the double Flip-Flop algorithm for parameter estimation of a tensor mixed effect model. For convergence, we test several convergence indices that are the divided $L_1$ norm of the difference between covariance matrices in two successive iterations, including $\left\|\widehat{\boldsymbol{\Sigma}}_i^{\{k\}} - \widehat{\boldsymbol{\Sigma}}_i^{\{k-1\}}\right\|_1 / J \cdot J, \left\|\widehat{\boldsymbol{\Psi}}_i^{\{k\}} - \widehat{\boldsymbol{\Psi}}_i^{\{k-1\}}\right\|_1 / K \cdot K, \left\|\widehat{\boldsymbol{\Omega}}_i^{\{k\}} - \widehat{\boldsymbol{\Omega}}_i^{\{k-1\}}\right\|_1 / L \cdot L$ for the first loop and $\left\|\widehat{\boldsymbol{\Sigma}}_\varepsilon^{\{t\}} - \widehat{\boldsymbol{\Sigma}}_\varepsilon^{\{t-1\}}\right\|_1 / J \cdot J, \left\|\widehat{\boldsymbol{\Psi}}_\varepsilon^{\{t\}} - \widehat{\boldsymbol{\Psi}}_\varepsilon^{\{t-1\}}\right\|_1 / K \cdot K, \left\|\widehat{\boldsymbol{\Omega}}_\varepsilon^{\{t\}} - \widehat{\boldsymbol{\Omega}}_\varepsilon^{\{t-1\}}\right\|_1 / L \cdot L$ for the second loop. We can see the convergence indices versus iterative histories in Fig. 3. We can find that the convergence history is monotonic and fast.

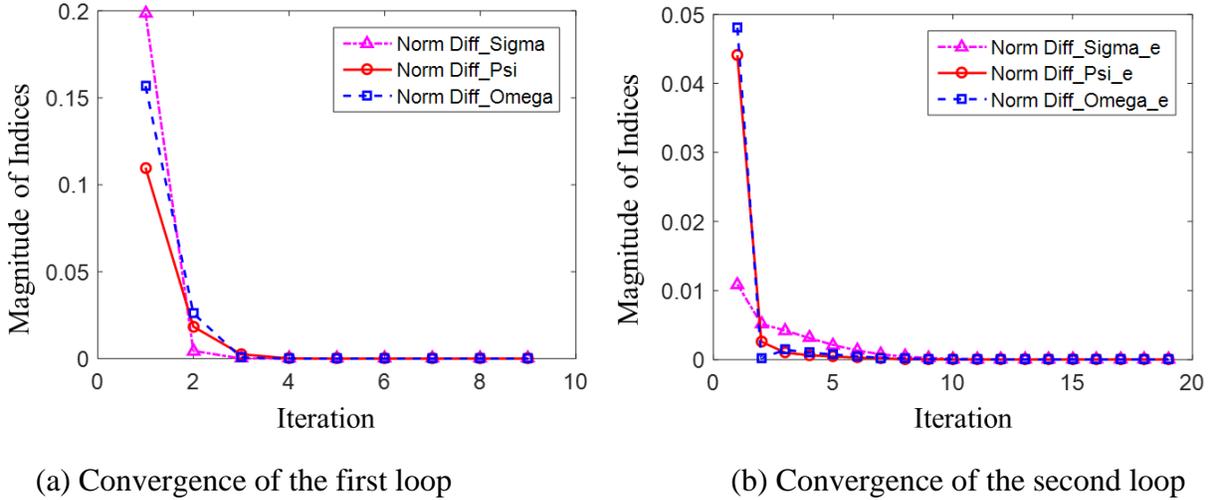

(a) Convergence of the first loop  (b) Convergence of the second loop

Fig. 3. Convergence of the iterative algorithm

When we do the parameter estimation for the TME model, we will get the design matrices $\boldsymbol{A}_i^{(1)}, \boldsymbol{A}_i^{(2)}, \boldsymbol{A}_i^{(3)}$ first by a Tucker decomposition for the mean response tensor. The design matrices $\boldsymbol{B}_i^{(1)}, \boldsymbol{B}_i^{(2)}, \boldsymbol{B}_i^{(3)}$ are a subset of the design matrices $\boldsymbol{A}_i^{(1)}, \boldsymbol{A}_i^{(2)}, \boldsymbol{A}_i^{(3)}$. After convergence, we compare the estimated parameters with a sample size of 600 and the ones in the simulation model (underlying true parameters). The parameters that we consider include core tensor of fixed effects, total covariance matrices from the first loop and covariance matrices of residual errors from the second loop. The results are shown in Fig. 4. We can see the estimations are quite consistent with the simulated parameters (underlying true parameters).

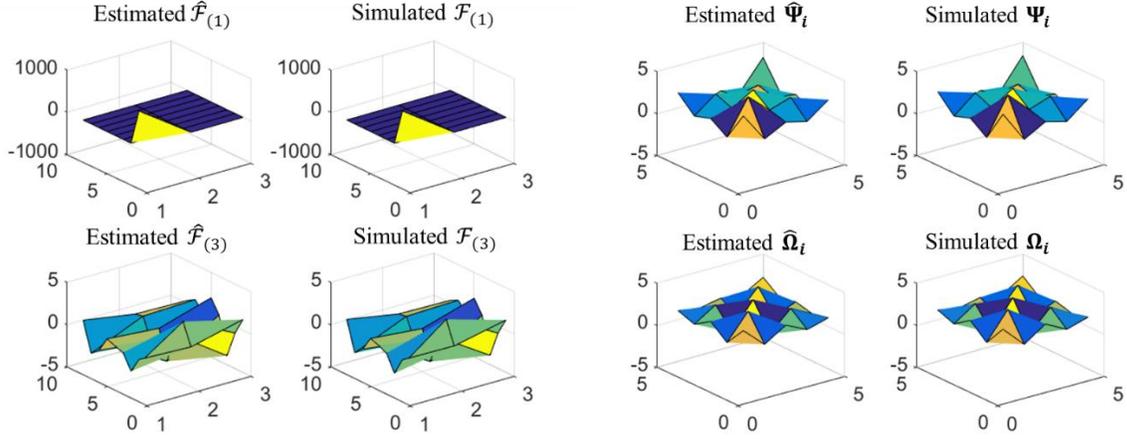

(a) Core tensor of fixed effects  (b) Covariance matrices $\widehat{\boldsymbol{\Psi}}_i$ and $\widehat{\boldsymbol{\Omega}}_i$

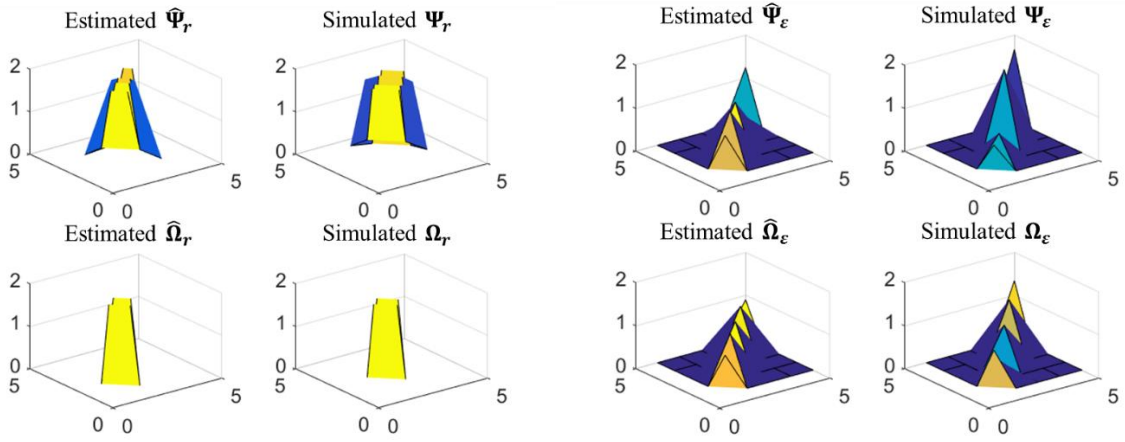

(c) Covariance matrices $\widehat{\boldsymbol{\Psi}}_r$ and $\widehat{\boldsymbol{\Omega}}_r$  (d) Covariance matrices $\widehat{\boldsymbol{\Psi}}_\varepsilon$ and $\widehat{\boldsymbol{\Omega}}_\varepsilon$

Fig. 4. Comparison between estimated parameters and simulated parameters

In order to quantitatively evaluate the estimation accuracy, we introduce indices, including $D_{\mathcal{F}}$, $D_{\boldsymbol{\Sigma}_i}$, $D_{\boldsymbol{\Psi}_i}$, $D_{\boldsymbol{\Omega}_i}$ for the first loop, and $D_{\boldsymbol{\Sigma}_\varepsilon}$, $D_{\boldsymbol{\Psi}_\varepsilon}$, $D_{\boldsymbol{\Omega}_\varepsilon}$ for the second loop. Where $D_X = \|\widehat{X} - X\|_F / \|X\|_F$, and $\widehat{X} - X$ denote the difference between estimated and true matrix/tensor. Moreover, $X = \{\mathcal{F}, \boldsymbol{\Sigma}_i, \boldsymbol{\Psi}_i, \boldsymbol{\Omega}_i, \boldsymbol{\Sigma}_\varepsilon, \boldsymbol{\Psi}_\varepsilon, \boldsymbol{\Omega}_\varepsilon\}$, and $\|\cdot\|_F$ denotes Frobenius norm. Furthermore, we introduce three indices for showing the convergence speed of different sample sizes. These indices are the iteration number, the time per iteration in the first loop, and the time per iteration in the second loop.

In order to explore the quantitative estimation accuracy and the asymptotic properties, we conduct the parameter estimation of the TME model for different sample sizes from 50 to 800. One hundred simulation runs are tested, where the mean and the standard deviation of the quantitative indices $D_X, X = \{\mathcal{F}, \Sigma_i, \Psi_i, \Omega_i, \Sigma_\varepsilon, \Psi_\varepsilon, \Omega_\varepsilon\}$ are calculated. The results are listed in Table 2. For the convergence speed, we notice that as the increase of sample size from 50 to 800, the average iteration number becomes smaller (from 9.32 to 6.05). While the average time per iteration increases from 0.62 seconds to 9.89 seconds in the first loop and from 0.22 seconds to 3.48 seconds in the second loop. For the quantitative estimation accuracy, the indices $D_\mathcal{F}, D_{\Sigma_i}, D_{\Psi_i}, D_{\Omega_i}, D_{\Sigma_\varepsilon}, D_{\Psi_\varepsilon}, D_{\Omega_\varepsilon}$ become smaller as the sample size increases. Which means that the estimated parameters are more accurate for the larger sample size. Of course, it costs more to obtain larger size of samples.

Table 2. Quantitative results for convergence speed and accuracy for different sample size

| Sample size | Iteration number | $D_\mathcal{F}$ | $D_{\Sigma_i}$ | $D_{\Psi_i}$ | $D_{\Omega_i}$ | Time 1/s | $D_{\Sigma_\varepsilon}$ | $D_{\Psi_\varepsilon}$ | $D_{\Omega_\varepsilon}$ | Time 2/s |
|---|---|---|---|---|---|---|---|---|---|---|
| 50 | 9.21 (0.48) | 0.0093 (0.0013) | 0.9463 (0.0016) | 0.0999 (0.0131) | 0.0353 (0.0151) | 0.63 (0.06) | 0.9596 (0.3692) | 0.2900 (0.0298) | 0.3586 (0.0344) | 0.27 (0.03) |
| 80 | 8.69 (0.46) | 0.0071 (0.0009) | 0.9140 (0.0022) | 0.0988 (0.0128) | 0.0325 (0.0161) | 0.99 (0.11) | 0.5566 (0.2811) | 0.2685 (0.0234) | 0.3327 (0.0269) | 0.41 (0.05) |
| 100 | 8.12 (0.48) | 0.0063 (0.0008) | 0.8924 (0.0027) | 0.0989 (0.0129) | 0.0318 (0.0158) | 1.26 (0.23) | 0.3966 (0.1835) | 0.2577 (0.0215) | 0.3188 (0.0240) | 0.49 (0.06) |
| 200 | 7.02 (0.14) | 0.0045 (0.0006) | 0.7842 (0.0049) | 0.0985 (0.0121) | 0.0297 (0.0157) | 2.47 (0.27) | 0.1827 (0.0355) | 0.2313 (0.0183) | 0.2876 (0.0191) | 0.99 (0.13) |
| 400 | 6.71 (0.45) | 0.0033 (0.0005) | 0.5682 (0.0094) | 0.0981 (0.0122) | 0.0280 (0.0158) | 4.94 (0.50) | 0.1249 (0.0205) | 0.2189 (0.0166) | 0.2736 (0.0180) | 1.99 (0.26) |
| 600 | 6.27 (0.45) | 0.0027 (0.0004) | 0.3525 (0.0143) | 0.0982 (0.0124) | 0.0277 (0.0162) | 7.42 (0.77) | 0.1100 (0.0182) | 0.2157 (0.0165) | 0.2696 (0.0178) | 2.97 (0.34) |
| 800 | 6.07 (0.26) | 0.0023 (0.0003) | 0.1380 (0.0181) | 0.0982 (0.0126) | 0.0277 (0.0162) | 9.86 (1.05) | 0.1033 (0.0172) | 0.2135 (0.0164) | 0.2678 (0.0175) | 3.98 (0.46) |

We compare our proposed TME method with two benchmark methods. The first is the tensor normal model with a structured mean (Nzabanita et al. 2015), which corresponds to the model that only considers fixed effects. We name it as the Tensor Fixed Effects (TFE) model and it is shown in Equation (18).

$$\mathcal{Y}_i = \mathcal{F} \times_1 A_i^{(1)} \times_2 A_i^{(2)} \times_3 A_i^{(3)} + \mathcal{E}_i \tag{18}$$

where the distribution of the residual errors tensor $\mathcal{E}_i$ is $N_{J,K,L}(\mathcal{O}; \Sigma_\varepsilon, \Psi_\varepsilon, \Omega_\varepsilon)$, and the noise covariance matrices along different dimensions are $\Sigma_\varepsilon \in \mathbb{R}^{J \times J}, \Psi_\varepsilon \in \mathbb{R}^{K \times K}, \Omega_\varepsilon \in \mathbb{R}^{L \times L}$. The second benchmark method is the Tucker decomposition (TD) for the average tensor response. In this method, we do not consider that the residual errors follow the tensor normal distribution.

In the simulation, we used the tensor toolbox from the Sandia National Laboratories (Bader, et al. 2015) when writing codes for the TME, the TFE and the TD. In order to evaluate the performance of the proposed TME model and benchmark methods, we use the mean square error for each sample denoted as $\text{MSE}_i = \|\boldsymbol{y}_i - \widehat{\boldsymbol{y}}_i\|_F^2 / JKL$ with $i = 1, \cdots, N$. The mean and standard deviation for MSE are calculated for different sample sizes and presented in Table 3. The time in Table 3 denotes the total running time of the corresponding model.

Table. 3. Comparison of mean square error in the TME model and benchmark methods

| Sample size | TME | | Benchmark 1 (TFE) | | Benchmark 2 (TD) | |
|---|---|---|---|---|---|---|
| | MSE | Time | MSE | Time | MSE | Time |
| 50 | 67.9450 (38.2969) | 9.29 | 12.4738 (1.2700) | 5.97 | 12.3996 (1.2719) | 1.69 |
| 80 | 26.2507 (9.4062) | 15.24 | 12.4155 (1.2760) | 8.20 | 12.3648 (1.2700) | 2.44 |
| 100 | 18.3721 (4.1957) | 18.08 | 12.3916 (1.2975) | 11.10 | 12.3483 (1.2960) | 2.86 |
| 200 | 11.9734 (1.2894) | 22.82 | 12.3624 (1.3992) | 17.37 | 12.3367 (1.4047) | 5.75 |
| 400 | 10.5248 (0.9193) | 36.95 | 12.3620 (1.4640) | 29.57 | 12.3478 (1.4666) | 11.80 |
| 600 | 10.2208 (0.8512) | 58.13 | 12.3860 (1.4300) | 46.93 | 12.3772 (1.4315) | 18.35 |
| 800 | 10.0595 (0.8234) | 85.55 | 12.3226 (1.4232) | 70.63 | 12.3167 (1.4235) | 26.06 |

The results of mean and standard deviation for MSE and computational time in different sample sizes are shown in Table 3. For the general pattern, as the sample size increases, the mean of MSE tends to become smaller for all those methods. This is because the quantitative estimation accuracy is low when the sample size is small. When the sample size is 50, 80, or 100, the MSE of the proposed TME model is larger than that of the TFE model and the TD model. The reason is that the sample size is lower than the number of unknown parameters needed to be estimated, which is 152 in this simulation example. Therefore the parameter estimation is not accurate. It indicates that if the sample size is low, the error from parameter estimation will significantly reduce the effectiveness of the model that considers random effects. Hence, it is better to only consider the fixed effects. When the sample size is larger than 200, the proposed TME model outperforms the TFE model and the TD model with respect to MSE. This is especially true when the sample size is comparable or larger than the total dimensions ($J \cdot K \cdot L$ in this example). The reason is that the TME model considers not only the fixed effects, but also the random effects. Additionally, the computational time of the TME model, the TFE model, and the TD model are comparable.

We also tried to conduct the vectorized Linear Mixed Effects (vLME) model, which conducts the typical linear mixed effects model (Galecki and Burzykowski 2013) after vectorization of the tensor responses. We used the fitlmematrix function in Matlab to conduct the vLME model. In this simulation, we transform $\boldsymbol{\mathcal{Y}} = \{\boldsymbol{\mathcal{Y}}_1, \boldsymbol{\mathcal{Y}}_2, \ldots, \boldsymbol{\mathcal{Y}}_N\}$ into a matrix, and get basis matrices from the vectorized parameters in $\{\boldsymbol{A}_i^{(1)}, \boldsymbol{A}_i^{(2)}, \boldsymbol{A}_i^{(3)}, \boldsymbol{B}_i^{(1)}, \boldsymbol{B}_i^{(2)}, \boldsymbol{B}_i^{(3)}\}$. However, vectorization destroys the tensor structure and results that the corresponding basis matrices are not of full rank. Therefore the vLME model failed in this situation. If we use different basis design in the TME and vLME models, the comparison will be unfair. Therefore, the TME model and vLME cannot be compared in the simulation.

It is worth mentioning that the small MSE is to show the TME model can realize accurate parameter estimation when there exist fixed effects and random effects in the multi-dimensional arrays. The main reason that we recommend the TME model is not that it can realize smaller MSE, but it can (i) separate fixed effects and random effects in a tensor domain; (ii) explore the correlations along different dimensions. If we know that there are not random effects in the tensor datasets according to domain knowledge, the TFE model and the TD model are recommended.

## 5.2 Surrogated Data Analysis of Raman Mapping

In this section, the performance of the TME model is evaluated through the surrogated Raman mapping data from a real CNTs buckypaper fabrication process. The setup of in-line Raman spectroscopy is shown in Fig. 5. In the experimental setup, Near Infra-Red (NIR) laser with a wavelength of 785nm and a laser output power of 150mW were used to eliminate the effect of ambient light. A low magnification lens was used to achieve a larger focus tolerance.

Raman mapping data have been collected from multiple rectangular zones, and the Raman data from each zone corresponds to one tensor. One Raman mapping tensor is shown in Fig. 2. Red dots represent measurement points, and there is a Raman spectrum in each measurement point. The mean response tensor is computed, and Tucker decomposition is conducted to obtain the design matrices. The dimension of the response tensor is 256×5×5. The dimensions of core tensor of fixed effects and random effects are 8×3×3 and 4×2×2, respectively. The covariance matrices of random effects are generated by weighted summation of diagonal matrices with random values and identity matrices. Two covariance matrices of residual errors are generated by the identity matrices with dimension 5×5 and another covariance matrix with dimension 256×256 that is diagonal with a given signal-dependent noise from experimental data.

After generating the surrogated Raman mapping data, the proposed TME is applied to extract different components including fixed effects, random effects, and signal-dependent noise. To explore the quantitative estimation accuracy, the same indices $D_{f1}, D_{\Sigma_i}, D_{\Psi_i}, D_{\Omega_i}, D_{\Sigma_\varepsilon}, D_{\Psi_\varepsilon}, D_{\Omega_\varepsilon}$ that were defined in Section 5 have been used to evaluate the results under different sample sizes. The results are shown in Table 4.

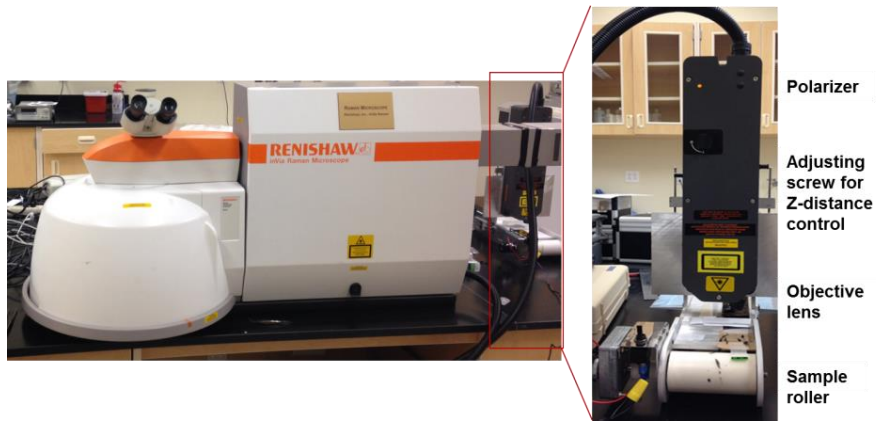

Fig. 5. Renishaw™ inVia micro-Raman system with custom-designed remote optical probe and roller sample stage

Table. 4. Quantitative results for convergence speed and accuracy for surrogated Raman mapping data

| Sample size | Iteration number | $D_{\mathcal{F}}$ | $D_{\Sigma_i}$ | $D_{\Psi_i}$ | $D_{\Omega_i}$ | Time 1/s | $D_{\Sigma_\varepsilon}$ | $D_{\Psi_\varepsilon}$ | $D_{\Omega_\varepsilon}$ | Time 2/s |
|---|---|---|---|---|---|---|---|---|---|---|
| 50 | 17 | 0.0039 | 0.9184 | 0.2031 | 0.2076 | 7.08 | 4.6351 | 0.3624 | 0.2261 | 3.55 |
| 80 | 13 | 0.0035 | 0.8693 | 0.2063 | 0.2158 | 11.94 | 4.3259 | 0.3455 | 0.2165 | 5.55 |
| 100 | 13 | 0.0031 | 0.8365 | 0.2079 | 0.2174 | 15.83 | 3.9043 | 0.3276 | 0.2087 | 6.91 |
| 200 | 9 | 0.0022 | 0.6735 | 0.2089 | 0.2190 | 30.84 | 0.2483 | 0.2945 | 0.1815 | 13.93 |
| 400 | 9 | 0.0015 | 0.3465 | 0.2103 | 0.2201 | 56.83 | 0.1449 | 0.2900 | 0.1792 | 28.62 |
| 600 | 7 | 0.0012 | 0.0702 | 0.2107 | 0.2206 | 90.73 | 0.1264 | 0.2908 | 0.1789 | 42.32 |

In comparison to Table 2, the results in Table 4 show similar asymptotic patterns, which means the estimated parameters become more accurate as the sample size increases. For example, when the sample size is 600, the indices $D_{\mathcal{F}}, D_{\Sigma_i}, D_{\Psi_i}, D_{\Omega_i}$ are as low as 0.0012, 0.0702, 0.2107 and 0.2206, respectively, which indicates the accuracy of the parameter estimation. However, the iteration number and time per iteration become larger for the same sample size due to the dimension increases from 30 to 256. Specifically, the computation time for 600 samples are 90.73 seconds per first loop and 42.32 per second loop.

## 5.3 Real Case Study

In this section, we show a real case study of applying the proposed TME model. The setup of in-line Raman spectroscopy is shown in Fig. 5. Similar to the surrogated data analysis, NIR laser with a wavelength of 785nm and a laser output power of 150mW were used to eliminate the effect of ambient light. The multi-walled CNTs buckypaper before alignment and after alignment are measured by Raman mapping technique. The scanning electron microscope (SEM) pictures for CNTs buckypaper are shown in Fig. 6. Alignment was conducted by stretching with different stretch ratio, including 0%, 20%, 35%, and 60%. When stretch ratio equals to 0%, it is referring to the CNTs buckypaper without alignment. The matrices $\Psi_r$ and $\Psi_\varepsilon$ are associated with the correlation along the horizontal direction, while the matrices $\Omega_r$ and $\Omega_\varepsilon$ are associated with the correlation along the vertical direction.

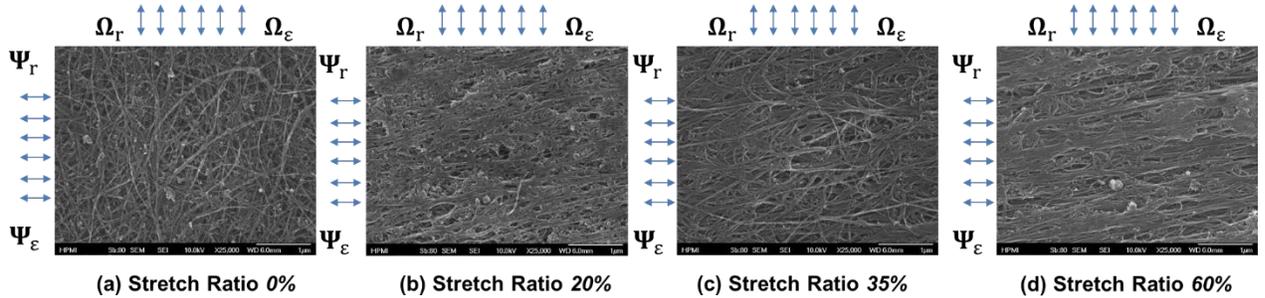

(a) Stretch Ratio *0%*  (b) Stretch Ratio *20%*  (c) Stretch Ratio *35%*  (d) Stretch Ratio *60%*

Fig. 6. SEM pictures of CNTs buckypaper with different stretch ratios

After running Raman mapping in a rectangular zone, 800 response tensors with dimension 256×3×4 are generated from each CNTs buckypaper sample. The proposed TME model is used to fit the datasets and the double Flip-Flop algorithm is conducted for parameter estimation. The covariance matrices $\Psi_r$, $\Psi_\varepsilon$, $\Omega_r$ and $\Omega_\varepsilon$ for CNTs buckypaper with different degrees of alignment are summarized in Fig. 7 and Fig. 8. Fig. 7(a) shows the range of diagonal entries in covariance matrices ($\mathbf{\Psi}_r$ and $\mathbf{\Psi}_\varepsilon$) along the horizontal direction; while Fig. 7(b) shows the range of diagonal entries in covariance matrices ($\mathbf{\Omega}_r$ and $\mathbf{\Omega}_\varepsilon$) along the vertical direction. Fig. 8 indicates changes of covariance coefficients as the stretch ratio increases. Fig. 8(a) shows coefficients in $\mathbf{\Psi}_r$, and Fig. 8(b) shows coefficients in $\mathbf{\Omega}_r$.

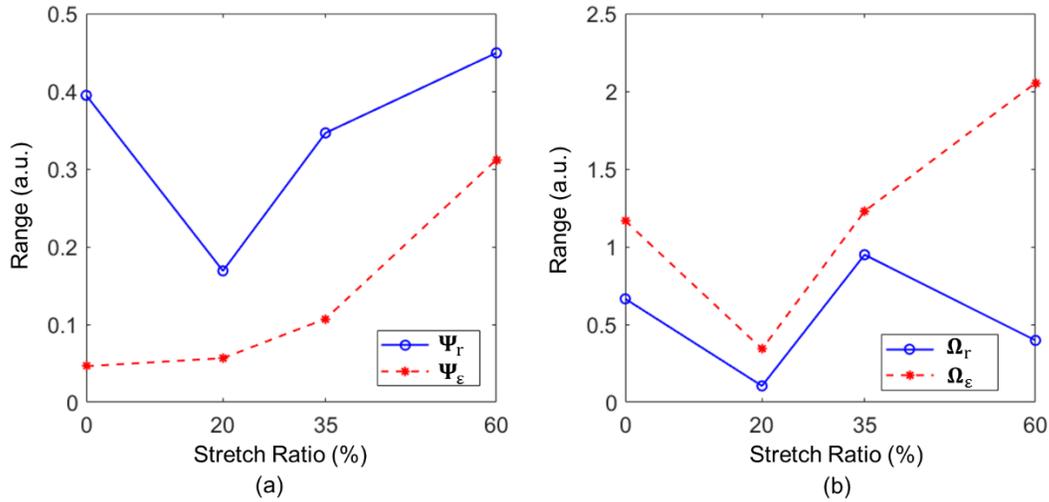

Fig. 7. Range of diagonal entries in covariance matrices (a) $\mathbf{\Psi}_r$ and $\mathbf{\Psi}_\varepsilon$, (b) $\mathbf{\Omega}_r$ and $\mathbf{\Omega}_\varepsilon$ for CNTs buckypaper with different degrees of alignment

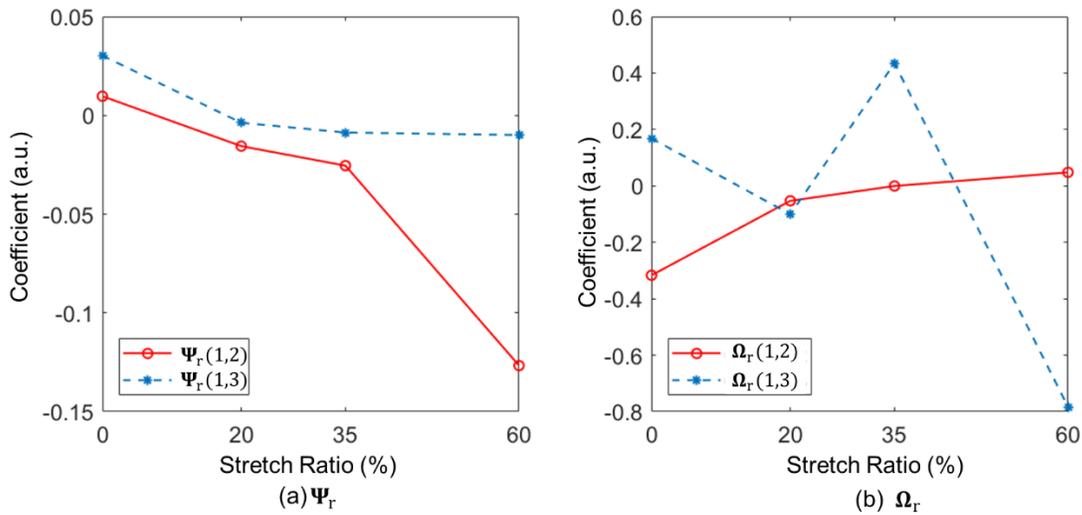

Fig. 8. Covariance coefficients in covariance matrices (a) $\mathbf{\Psi}_r$, (b) $\mathbf{\Omega}_r$ for CNTs buckypaper with different degrees of alignment

We compare the covariance matrices along horizontal and vertical directions for each CNTs buckypaper sample. Considering the physical knowledge of CNTs buckypaper, we can provide the following remarks:

- For the random effects covariance along the horizontal direction, we observe that the coefficient $\Psi_r(1,2)$ tends to become negative after alignment and as the degree of alignment increases, the absolute magnitude becomes larger. The covariance coefficient $\Psi_r(1,2)$ changes from 0.0097 to -0.1267. It indicates that negative correlation along the alignment direction occurs, and the covariance coefficient changes with the alignment of CNTs buckypaper. This can be explained by the conservation of mass in a local zone. Alignment introduces systematic ridges and valleys in the microstructure pattern, and a ridge will be close to a valley, that indicates the negative correlation in the height. The height will impact the measurement distance between the laser head in Raman spectroscopy and the Raman mapping. The covariance coefficient in $\Psi_r(1,3)$ becomes negative after alignment, but the absolute magnitude becomes closer to zero. One physical interpretation is that after alignment, the distance between the first measurement line and the third measurement line becomes larger, and their correlation relationship becomes weaker.

- For the random effects covariance along the vertical direction, coefficients $\Omega_r(1,2)$ becomes closer to zero, but the absolute magnitude of $\Omega_r(1,3)$ becomes larger. The physical interpretation is that as the stretch ratio increases, the high-frequency surface roughness becomes smaller, while the low-frequency surface roughness becomes larger.

- For the range of diagonal entries in covariance matrices of Raman mapping, the change of $\Psi_r$, $\Omega_r$ and $\Omega_\varepsilon$ for CNTs buckypaper with different degrees of alignment are quite random. However, $\Psi_\varepsilon$ has a larger quantitative difference between the maximum entry and the minimum one after alignment. Without alignment, the diagonal coefficients range is 0.0464, while after alignment with a stretch ratio of 60%, the range becomes as large as 0.3121. Which means that for different measurement lines along the alignment direction, the variability becomes larger. This makes sense because the alignment creates systematic ridges and valleys along the alignment directions. We can use this index to quantify the degree of alignment.

In summary, based on the covariance matrices from the TME model, we can quantify the influence of alignment based on the range of diagonal entries in $\mathbf{\Psi}_\varepsilon$ and covariance coefficients $\mathbf{\Psi}_r(1,2)$. The quantitative changes after alignment can be interpreted by engineering knowledge. We want to point out that this case study is chosen to illustrate the performance of our approach. Other approaches may also work well for quantifying the degree of alignment. In addition, our TME approach can be extendable to other applications.

## 6. Summary

In this paper, we proposed a novel TME model that effectively and efficiently explores the fixed effects and random effects inherent to the data in tensor domain. The advantages of this model include (i) its capability to handle multilevel hierarchical data; (ii) its ability to take complexed association structures, including correlation along different dimensions, into consideration; (iii) analyzing the mixed effects for the high-dimensional datasets. The proposed TME model can be viewed as a logical extension from a vector/matrix-valued mixed effects model to an array-valued mixed effects model. The proposed TME model is applied in the nanomanufacturing inspection. Moreover, the TME model can be applied to provide potential solutions for a family of tensor data analytics with mixed effects, such as problems in the research fields of multimodality imaging analysis, chemometrics, neuroimaging, multichannel signal processing, etc.

For the TME model, the distribution of response tensors and its $k$-mode matricization were explored. We also derived the log-likelihood function for the TME model. Maximum likelihood estimators for fixed effect core tensor and covariance matrices were derived. Existence of the MLE and identifiability of the TME model were illustrated. Moreover, an iterative double Flip-Flop algorithm has been developed for parameter estimation, and the initialization and convergence criteria have been discussed. The computational complexity of the Flip-Flop algorithm has been derived. The TME model was shown to outperform vectorized LME model from a computational complexity perspective. By simulation and surrogated data analysis, we found that the algorithm can realize very quick convergence. The iteration number becomes smaller and time per iteration becomes longer as the sample size increases. In addition, the asymptotic property was investigated in the simulation and surrogate data analysis. The estimation accuracy of total covariance matrices and covariance matrices for the error terms improve as the sample size increases. In the simulation study, we also show that the TME model outperforms two benchmark methods which do not consider random effects (the TFE model and the TD model) when the sample size is larger than the dimensions of response tensor. Furthermore, in the case study, the influence of alignment of CNTs buckypaper is quantified by the covariance matrices along different dimensions.

In future work, different extracted components for Raman mapping of different kinds of buckypaper can be analyzed based on the TME model. The extracted features will be used to do inspection and monitoring of various quality characteristics, such as fabrication consistency, thickness variability, uniformity, and defect information. Finally, a set of quality assessment criteria of CNTs buckypaper will be developed.

**Supplementary Materials**

Supplementary materials contain the data and functions for the TME model and all technical proofs.